\documentclass{iaus}
\usepackage{graphicx}

\def\kms{~km~s$^{-1}$}
\def\h2o{H$_2$O}
\def\vlsr{$V_{\mbox{\scriptsize LSR}}$\ }
\def\etal{ et~al.\ }
\def\arcdeg{$^{\circ}$\hspace{-2pt}}

\def\as1828{IRAS~18286$-$0959}
\def\ras1846{IRAS~18460$-$0151}
\def\iras1913{IRAS~19134$+$2131}

\title[Stellar molecular jets traced by maser emission]
{Stellar molecular jets traced by maser emission}

\author[H.~Imai]
{Hiroshi Imai}

\affiliation{Department of Physics, Faculty of Science, Kagoshima University, 
Japan \break email: hiroimai@sci.kagoshima-u.ac.jp}

\pubyear{2007}
\volume{242}  
\pagerange{000--00}
\date{16 April 2007}
\jname{Proceedings Title IAU Symposium}
\editors{J.~Chapman \& W.~Baan, eds.}

\begin{document}
\maketitle

\begin{abstract}
Highly collimated jets found in AGB and post-AGB stars are expected to play an important role for shaping planetary nebulae. Recent VLBI observations of \h2o  maser sources have revealed that some of the spatio-kinematical structures of \h2o maser sources exhibit stellar jets with extremely spatially and kinematically high collimation. Such stellar \h2o maser jets, so-called  "water fountain" sources, have been identified in about 10 sources to date. Here we review recent VLBI observations and relevant observational results of the water fountain sources. They have revealed a typical dynamical age and the detailed kinematical structures of the water fountains, possibility of the existence of "equatorial flows", and the evolutionary status of the host stars. The location and kinematics of one of the water fountain sources in the Galaxy is also revealed and shown here. 

\keywords{masers, stars:AGB and post-AGB, mass loss, winds, outflows}
\end{abstract} 

\section{Introduction}
A stellar jet appearing at the final stage of stellar evolution is a manifestation of the important phenomena to form asymmetric planetary nebulae (PNe). Based on the large incidence of bipolar and multipolar optical morphologies and the common presence of point-symmetry found in a sample of young PNe, \cite{sah98} concluded that collimated jets have been operational in a large fraction of PNe, and expected to be one of the major factors shaping these objects. Similar morphologies have been found in a survey of pre-PNe (\cite{sah04}), leading to the current paradigm that such a collimated jet had already been launched at an earlier phase of stellar evolution  -- the asymptotic giant branch (AGB) phase - and is frequently observable in molecular line emission. Subsequently an optically visible PN is seen at the post-AGB phase.

There are candidate sources that show us the earliest stage of stellar jet emergence. ``Water fountains" are the most promising candidates; they exhibit extremely-high-velocity flows traced by \h2o\ maser emission. The outflow velocity is much faster than the typical expansion velocity of circumstellar envelopes of AGB stars found in 1612~MHz OH masers (10--25\kms, e.g., \cite{lin89}). The \h2o maser emission exhibits high collimation in its both morphology and kinematics. This suggests that high density gas is supplied from the vicinity of the stellar surface to the jet tips where \h2o\ maser emission is excited (e.g.,  \cite{eli89}). Because VLBI observations have been made for some of these water fountains, it is possible to undertake statistical analysis to estimate typical values of their physical parameters, lead to understanding the final evolution of the circumstellar envelopes and the formation mechanism of stellar jets.

Note that high precision astrometry of \h2o masers in the water fountains provides great opportunity to directly estimate the locations and kinematics of the water fountains in the Galaxy, leading to estimating masses and birth points of the stars driving these water fountains. 

\begin{table}[t]
\caption{Parameters of "water fountain" sources.}
\label{tab:water-fountains}
\begin{tabular}{llccccl}
\\ \hline \hline 
& & Length & $ \Delta V_{\rm los}$\footnotemark[1] 
& $t_{\rm jet}$\footnotemark[2] & Distance & \\
IRAS name & Other name & (arcsec) & (km~s$^{-1}$) & (year) & (kpc) 
& Reference\footnotemark[3] \\ \hline
15445$-$5449 & OH~326.5$-$0.4 & & 91 & & & 4 \\
15544$-$5332 & OH~325.8$-$0.3 & & 40 & & & 4 \\
16342$-$3814 & OH~344.1$+$5.8 & 2.4 & 240 & $ \sim$100 & 2 & 3, 15, 16, 17, 18 \\
16552$-$3050 & GLMP~498 & & 170 & & & 19 \\
18043$-$2116 & OH~0.9$-$0.4 & & 204 & & & 4 \\
18139$-$1816 & OH~12.8$-$0.9 & 0.11 & 55 & 70 & ? & 1, 2, 10 \\
18286$-$0959 & & 0.24 & 200 & $\sim$15 & 3.1 & 5, 11 \\
18450$-$0148 & W43A, OH31.0$+$0.0 & 0.8 & 190 & 50 & 2.6 & 6, 8, 13, 15,16, 20 \\
18460$-$0151 & OH~31.0$-$0.2 & 0.11 & 290 & $\sim$5 & 6.8 & 5, 11 \\
18596$+$0315 & OH~37.1$-$0.8 & & 59 & & & 4 \\
19134$+$2131 & & 0.15 & 100 & 40 & 8.0 & 12, 14, 16 \\ \hline
\end{tabular}

\noindent
\footnotemark[1]Full range of the line-of-sight velocities of maser emission. \\
\footnotemark[2]Dynamical age of the jet. \\
\footnotemark[3]1: \cite{bau79}, 2: \cite{bob05}, 3: \cite{cla04} 4: \cite{dea07}, 
5: \cite{deg07}, 6: \cite{dia85}, 7: \cite{eng02}, 8: \cite{gen77}, 9: \cite{eng86}, 10: \cite{gom94}, 11: Imai \etal\ 2007 in preparation, 12: \cite{ima07}, 
13: \cite{ima05}, 14: \cite{ima04}, 15] \cite{ima02}, 16: \cite{lik92}, 
17: \cite{mor03}, 18: \cite{sah99}, 19: \cite{sua07}, 20: \cite{vle06}
\end{table}

\section{Lifetime and timing of the water fountain}
Table \ref{tab:water-fountains} gives physical parameters of the water fountain jets derived from single-dish and interferometric observations. Lengths of stellar jets traced by \h2o maser emission range over 6000~AU (in IRAS~16342$-$3814, \cite{mor03,cla04}), which is still equal to or shorter than a radius of a circumstellar envelope of an AGB star (e.g., \cite{eli92}). On the other hand, flow velocities of them ($V_{\rm flow}\sim \Delta V_{\rm los}/2$) are sometimes over 100\kms. These give the dynamical ages of jets shorter than 100~years, only $ \sim$5~years in an extreme case (Imai \etal in preparation). In the time interval of 100~years, a stellar jet traveling with a typical water fountain velocity of 100\kms\ reaches $ \sim$2000~AU from the central star, which corresponds to the typical size of an OH maser shell in a circumstellar envelope around an OH/IR star, at the final stage of energetic mass loss. The \h2o maser emission is excited near the tip of the stellar jet where high compression takes place in the shock at the interface between the jet and the ambient circumstellar gas. But because the ambient gas density declines with radius, the maser emission may be quenched at the distance where the post-shock gas density falls below a critical value, $n_{\mbox{\tiny  H$_{2}$}}\sim$10$^6$ cm$^{-3}$ (\cite{eli89}). 

Note that while optical lobes are observed in some water fountains (IRAS~16342$-$3814, and \iras1913), the optical emission is not detected in others (W43A and OH12.8$-$0.9), suggesting that the presence of water fountains may help in the identification of new pre-PNe, which may be optically invisible due to heavy circumstellar extinction. The optical lobes in \iras1913 and IRAS~16342$-$3814 coexists with the \h2o maser emission (\cite{sah99,mor03, ima07}). 
It is expected that the lobes have been recently formed along the collimated fast jet, in which the circumstellar envelope around the evolved star has been penetrated by the jet and starlight. These sources may be at the similar stage in the transition from an AGB to post-AGB star.

Does the size of a circumstellar envelope determine the lifetime of a water fountain? 
We already know that \h2o masers can survive until photoionization of the circumstellar envelope starts to form a PN, such as K3$-$35  (\cite{mir01}). In K3$-$35, \h2o maser emission is seen at the tips of the radio lobes as well as in the inner and denser part of the circumstellar envelope where where UV radiation from the star has not yet penetrated. However, the maser emission at the lobe tips of K3-35 is only weakly collimated, both morphologically and kinematically, implying that an initially collimated jet traceable by \h2o maser emission is being destroyed. The dynamical age of the radio lobes is estimated to be 800~yr (\cite{mir01}), which may correspond to an upper limit to the lifetime of a water fountain source but is one order of magnitude longer than the dynamical age estimated using VLBI observations of the maser spatio-kinematical structures. 

Here we speculate that the collimated jet may drill through the circumstellar envelope formed during the AGB phase and the tip of the jet may reach the outer surface of the envelope before the onset of photoionization of the envelope. The masers lying beyond the optical lobes, such as those in IRAS~16342$-$3814, might soon be quenched within a few decades or centuries owing to the decline of the gas density in the pre-shock gas as the jet propagates outward, which in turn leads to a loss of collimation and ultimately to a loss of the maser inversion as the post-shock gas density declines.

\begin{figure}
    \includegraphics[width=13.5cm]{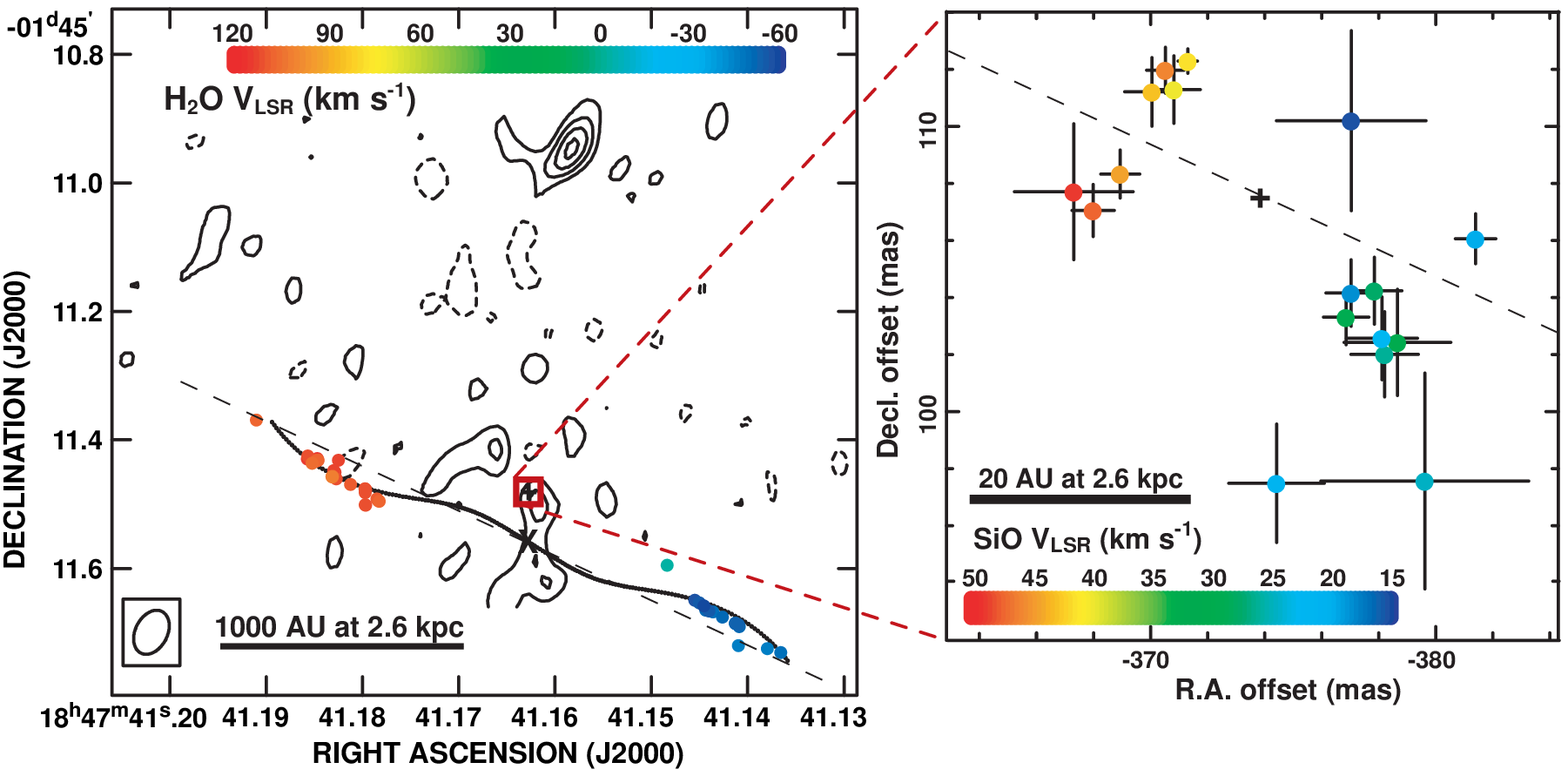}
\caption{\h2o, SiO ($v=$1, $J=$1--0) maser emission, and 7 mm continuum emission in W43A. Different colors denote maser features' line-of-sight velocities, according to the color scale. {\it Left} Spatio-kinematics of the \h2o\ maser features observed on 2002 April 3. A dotted line shows a pattern of a precessing jet appearing in the year 2002.3, which is modeled by \cite{ima02}. A symbol "X" indicates the dynamical center of the modeled jet. A thin dashed line indicates the axis direction of the \h2o\ maser jet. 
{\it Right} Spatio-kinematics of the SiO maser features. Horizontal and vertical bars 
of individual maser features indicate uncertainties of feature positions 
in R.A.\ and decl.\ directions, respectively. A plus symbol indicates a roughly-estimated location of the dynamical center of the biconically-expanding flow found in the SiO maser kinematics. Roughly a flux density of a spot is inversely proportional to an error bar of the spot.}
\label{fig:W43A-SiO}
\ \\
    \includegraphics[width=13.5cm]{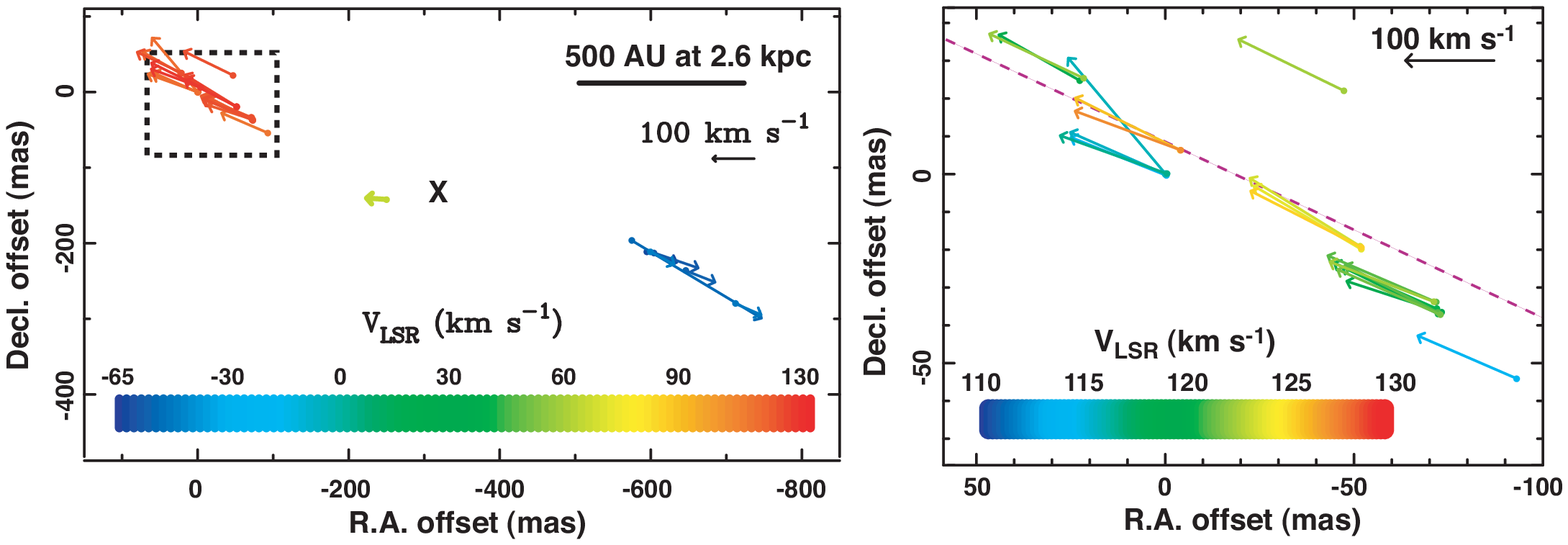}
\caption{Relative proper motions of \h2o masers in W43A measured during 2003--2004 with a mean proper motion subtracted. A cross indicates the location of the dynamical center of the precessing bipolar jet. {\it Left} the whole distribution of maser features. {\it Right} zoom-up view of the feature distribution in the area displayed in a dashed-line box in the left panel.}
\label{fig:W43A-PM}
\end{figure}

\section{Equatorial flows in the water fountain sources}

Highly collimated fast stellar jets as described above are expected to be driven by magneto-hydrodynammical process (e.g., \cite{bla01,vle06}). However it is obscure whether the stellar object driving the jet is a single star or a binary system; one is an AGB star providing material to another through an accretion disk. Astrometry for \h2o and 1612~MHz OH maser emission will provide an important clue to find each of the locations of the driving sources of the collimated jet and the AGB flow. Only W43A has SiO maser emission (\cite{nak03,ima05}), which provides a better opportunity of exactly find the location of the AGB star. 

Figure \ref{fig:W43A-SiO} compares the locations of the dynamical centers of the \h2o maser jet and the SiO maser flow using a common extragalactic position reference. Although astrometric accuracy is much poor in the north--south direction for W43A located very close to the celestial equator, the dynamical centers are coincide within 4~mas (10~AU at 2.6~kpc). The SiO maser flow is well modeled by a biconically expanding flow with a low velocity ($\sim$20\kms) and a large opening angle (\cite{ima05}). These implies that both of the collimated jet and the biconical flow are driven in the region within 10~AU. 

\begin{figure}
    \includegraphics[width=13.5cm]{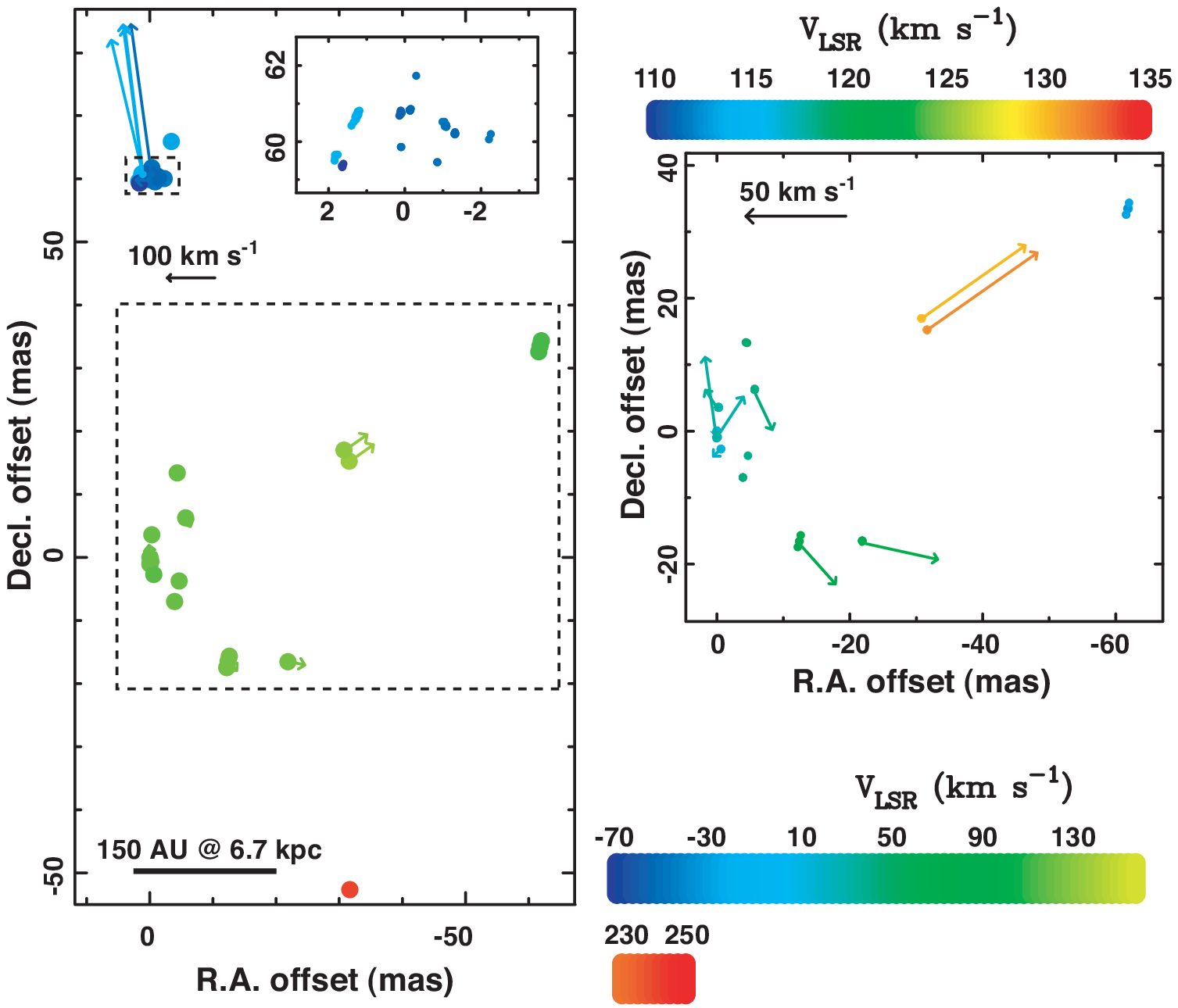}
    \includegraphics[width=13.5cm]{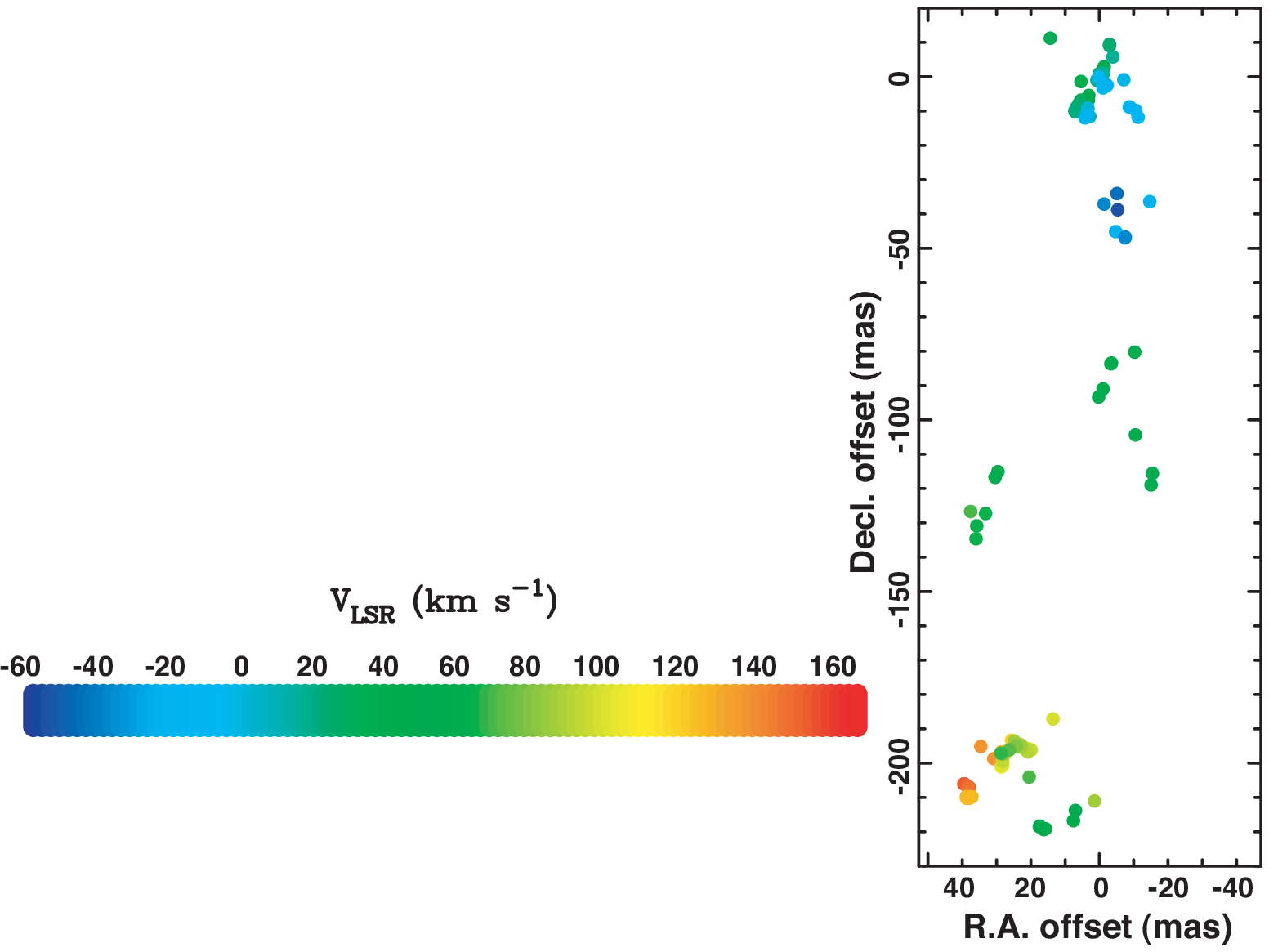}
\begin{minipage}{7.5cm}
\vspace*{-11cm}
\caption{Distribution (filled circle) and relative proper motions (arrows) of \h2o masers in \ras1846 during 2006 September--November. {\it Left} Overview of the distribution. {\it Right} Zoom-up view of the area displayed with a dashed line box in the left panel. Different colors denote the features' line-of-sight velocities, 
according to the color scale above this panel. {\it Relative} proper motions of maser features are measured with respect to the maser feature located at the origin and biased in the north--west direction by $ \sim$30\kms.}
\label{fig:IRAS1846}
\vspace{4.5cm}
\caption{Same as figure \ref{fig:IRAS1846}, but for \as1828 observed on 2006 September 14.}
\label{fig:IRAS1828}
\end{minipage}
\end{figure}

Note that the spatio-kinematical structure of the SiO masers does not exhibit any rotation or contraction as suggested when mass accretion occurs from a disk. Figure \ref{fig:W43A-PM} shows relative proper motions of the W43A \h2o masers. We found a maser feature that is located close to the dynamical center and has a Doppler velocity close to the systemic velocity of W43A (\vlsr$=$34\kms) and a proper motion corresponding to a speed of $\sim$30\kms. The velocity vector is pointed to the different direction from the jet major axis. The low velocity maser feature is expected to be associated with an "equatorial" flow, rather than an accretion disk.
Such equatorial flows have recently been recognized in other water fountain sources, 
\as1828 and \ras1846, as well (\cite{deg07}). Figure \ref{fig:IRAS1846} and \ref{fig:IRAS1828} shows the spatio-kinematics of \ras1846 and \as1828, respectively. The former source has the fastest jet ever observed with a speed over 300\kms. This source also has an equatorial flow with a velocity of $\sim$30\kms\ in the region close to the midpoint of the most blue-shifted and red-shifted maser features forming the high velocity jet. The latter source has many velocity maser components, only the most blue-shifted and red-shifted maser features are clearly separated each other and each of them is distributed in a small area. These imply the coexistence of a collimated fast flow and a slow outflow associated with either an AGB envelope or an equatorial flow driven by the fast flow as its sub-product. The existence of such equatorial flows may support a hypothesis that a single star can create both of fast collimated and slow equatorial flows when the star is collapsing into a white dwarf (H.~Washimi and S.~Miyaji, in private communication).

\section{Ballistic corkscrew jets}
\begin{figure}
\begin{center}
    \includegraphics[width=11cm]{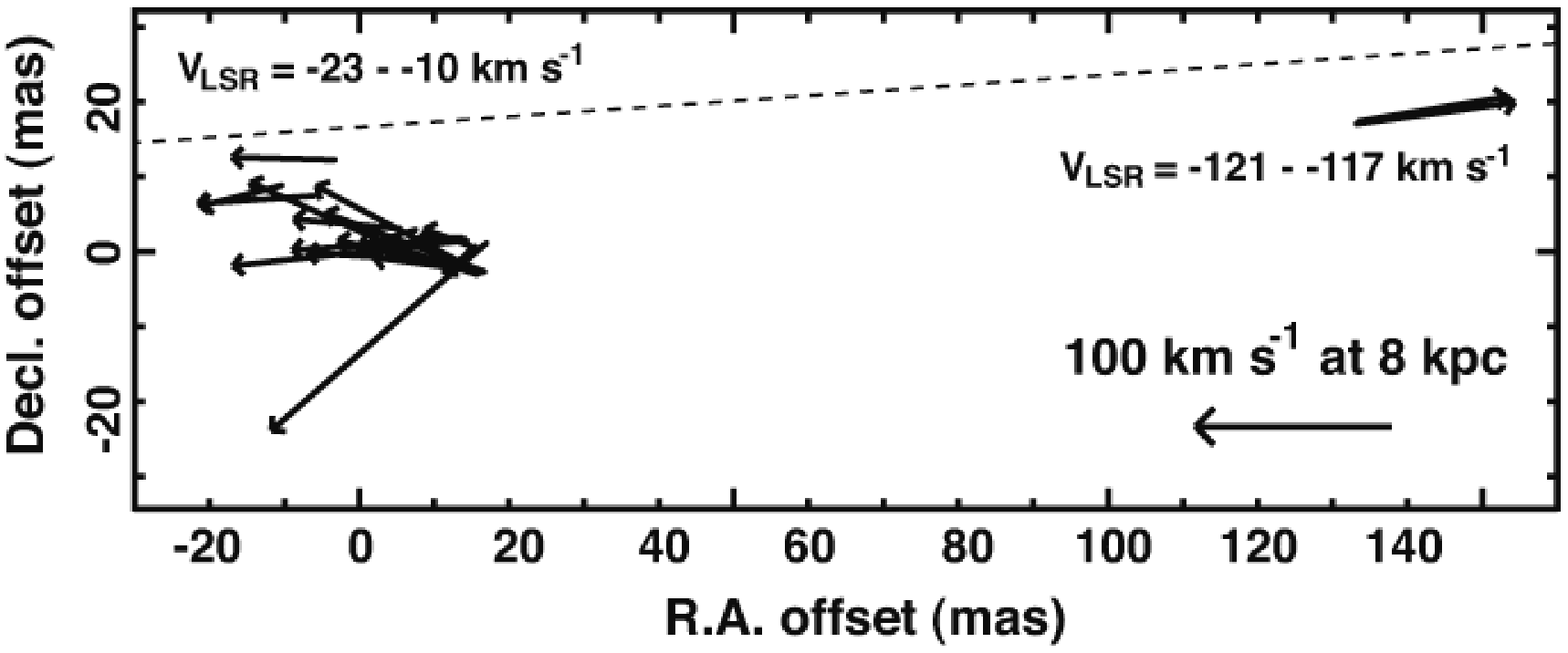}
\end{center}
\caption{\h2o maser feature distribution in \iras1913 (roots of arrows) and relative proper motion vectors of these masers (arrows) with respect to a position reference maser feature with a mean proper motion subtracted.}
\label{fig:I19134-PM}
\end{figure}

The morphology of water fountains may provide a unique clue to elucidating how to eject the material in the jets, leading to estimating the character of the central stars. Some of the water fountain jets exhibit spiral patterns. In particular, such a pattern is clearly seen in the distributions of \h2o masers in W43A (\cite{ima02,ima05}). On the other hand, a "corkscrew jet" is found in optical emission in IRAS~16342$-$3814 (\cite{sah05}). Recently we found twin spiral patterns of the \h2o maser distribution in W43A, one of the spiral is seen in the axisymetrically opposite side of another (dotted line) with respect to the jet major axis (thin dashed line) (see figure \ref{fig:W43A-SiO}. However, curved motions of the maser features are not recognized. The right panel of figure \ref{fig:W43A-PM} and figure \ref{fig:I19134-PM} show proper motions of \h2o masers found in W43A and \iras1913, respectively, in detail. In each source, we find an alignment of maser features whose direction has an offset from the major axis of the jet. On the other hand, proper motion vectors of maser features are still parallel to the jet axis. These suggest that the corkscrew jets found in these sources are formed by ballistic motions of maser features from the driving source of the jets, but whose directions are precessing with time. The alignments of maser features may be formed at the shock fronts interfaced with the ambient circumstellar envelopes. 

\section{Location and motion of the water fountain in the Galaxy}
\begin{figure}
    \includegraphics[width=13.5cm]{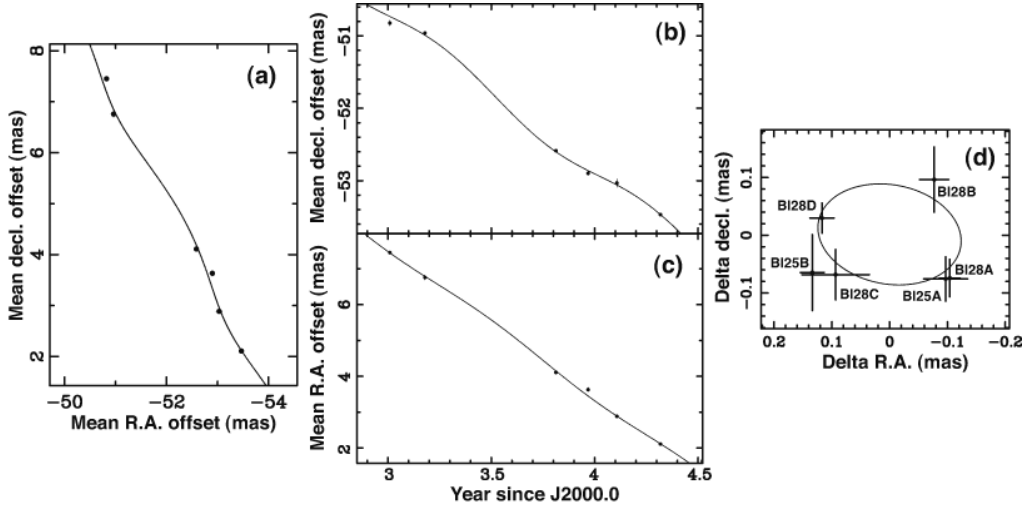}
\caption{Mean motion of three \h2o maser features in \iras1913 with respect to the extragalactic position reference source J1925$+$2106. ({\it a}) Mean R.A.\ and decl.\ offsets on the sky with respect to the phase-tracking center. ({\it b}) and ({\it c}) Mean R.A.\ and decl.\ offsets against time. ({\it d}) Relative mean offsets with a mean proper motion subtracted. The observation code is shown alongside each measured position.}
\label{fig:I19134-parallax}
\end{figure}

\h2o masers associated with \iras1913 (hereafter I19134) have measured their {\it absolute} coordinates with respect to the extragalactic position reference source J1925$+$2106 for a span over one year (\cite{ima07}). We therefore fit the temporal variation of the mean position of three maser features to a kinematical model consisting of an annual parallax and a constant velocity mean secular motion. Figure \ref{fig:I19134-parallax} shows the fitting result. We obtain a best-fit annual parallax, corresponding to a distance to I19134 $D=8.0^{+0.9}_{-0.7}$~kpc. On the basis of the inferred heliocentric distance and the systemic secular maser motion, we estimate the location of I19134 in the Galaxy to be, $(R, \theta, z)=(7.4^{+0.4}_{-0.3}$~kpc, 62\arcdeg$\pm$5\arcdeg, $0.65^{+0.07}_{-0.06}$~kpc,
and the 3-D velocity vector in Galactic cylindrical coordinates to be, 
$(V_{R}, V_{\theta}, V_{z})=
(3^{+53}_{-46}, 125^{+20}_{-28}, 8^{+48}_{-39})\mbox{ [km~s$^{-1}$]}$.

The estimated location is close to the Perseus spiral arm within $\sim$1~kpc (c.f., \cite{nak06}). Note that the estimated Galactic rotation velocity of I19134 ($\sim$125\kms) is much slower than that implied by the Galactic rotation curve ($\sim$220\kms, e.g., \cite{cle85, deh98}). It implies that the I19134's orbit must have an orbit with large excursions in radius although I19134 is orbiting along a circle at present. I19134 is apparently orbiting as a member of the Galactic "thick" component rather than the "thin" component. Adopting the source velocity perpendicular to the Galactic plane $V_{z}=$8\kms, we find that the present Galactic altitude gives a lower limit to the travel time of I19134 from the Galactic plane to the present position to be $\sim 2.4\times 10^{7}$~yr. This suggests that the stellar object of I19134 has a lifetime of at least this travel time and a stellar mass smaller than  $\sim$8.6~$M_{\odot}$ (here we adopt a stellar lifetime to be $t_{\ast}\propto M_{\ast}/L_{\ast}\propto M^{-2.8}_{\ast}$) if the central star of I19134 was born in the Galactic plane as a member of the Galactic "thin disk" component. The statistical study of Galactic PNe by \cite{man04} shows that the Galactic scale height of bipolar PNe such as I19134 is $<z>=$100~pc, suggesting that their progenitors should be high mass stars. Although the mass of I19134's progenitor is therefore expected to be near the upper mass limit estimated above, the present galactic altitude of I19134 and its anomalously slow galactic rotation velocity are quite odd. These may imply an abnormal environment during the birth of I19134's central star. However, the estimated mass of I19134 still has a big uncertainty 
due to the poor accuracy of the vertical velocity, $V_{z}$.

\ \\
\noindent{\bf Acknowledgements} 
The present works have been made in the combination of collaborations with S.~Deguchi, P.~J.~Diamond, S.~Kwok, A.~Miyazaki, M.~Morris, J.~Nakashima, K.~Obara, T.~Omodaka, R.~Sahai, T.~Sasao, and W.~H.~T.~Vlemmings. The NRAO's VLBA and VLA are facilities of the National Science Foundation, operated under a cooperative agreement by Associated Universities, Inc. The EVN is a joint facility of European, Chinese, South African and other radio astronomy institutes funded by their national research councils. H.I.\ was supported by IAU Grant for traveling to the symposium and Grant-in-Aid for Scientific Research from Japan Society for Promotion Science (18740109).

\end{document}